\documentclass{PoS}

\title{On isospin breaking in $\tau$ decays for $(g-2)_\mu$ from Lattice QCD}

\ShortTitle{On isospin breaking in $\tau$ decays for $(g-2)_\mu$ from Lattice QCD}

\author{\speaker{Mattia Bruno}\\
        Brookhaven National Laboratory\\
        E-mail: \email{mbruno@bnl.gov}}

\author{Taku Izubuchi\\
        Brookhaven National Laboratory}
\author{Christoph Lehner\\
        Brookhaven National Laboratory}
\author{Aaron S.~Meyer\\
        Brookhaven National Laboratory}

\abstract{
Hadronic spectral functions of $\tau$ decays have been used in 
the past to provide an alternative determination of the LO 
Hadronic Vacuum Polarization relevant for the (g-2) of the muon. 
Following recent developments and results in Lattice QCD+QED calculations, 
we explore the possibility of studying the isospin breaking corrections 
of $\tau$ spectral functions for this prediction. 
We present preliminary results at physical pion mass based on 
Domain Wall Fermion ensembles generated by the RBC/UKQCD collaboration. 
}

\FullConference{The 36th Annual International Symposium on Lattice Field Theory - LATTICE2018\\
		22-28 July, 2018\\
		Michigan State University, East Lansing, Michigan, USA.}

\newcommand{\up}{\mathrm u}
\newcommand{\down}{\mathrm d}

\newcommand{\alphaem}{\alpha_\mathrm{em}}
\newcommand{\FSR}{\mathrm{FSR}}
\newcommand{\GEM}{G_\mathrm{EM}}
\newcommand{\SEW}{S_\mathrm{EW}}
\newcommand{\RIB}{R_\mathrm{IB}}

\newcommand{\eq}[1]{eq.~(\ref{eq:#1})}
\newcommand{\fig}[1]{Fig.~(\ref{fig:#1})}
\newcommand{\Ref}[1]{Ref.~\cite{#1}}
\newcommand{\Refs}[1]{Refs.~\cite{#1}}

\begin{document}

\section{Introduction}

The discrepancy between the theoretical determination
of the muon anomalous magnetic moment $a_\mu$ and 
the experimental measurement performed at the Brookhaven
National Lab. is a very interesting place to look for 
new physics beyond the Standard Model. 
While new experimental results are expected from 
the Fermilab and J-PARC Laboratories, theoretical calculations
are trying to consolidate or improve their error estimates. 
A crucial piece is the hadronic
vacuum polarization (HVP) where data-driven approaches based
on dispersive analysis are presently overpowering 
pure first-principles non-perturbative calculations from the 
lattice.

Data-driven approaches are essentially based on 
 results from dedicated measurements of cross sections 
of $e^+ e^- \to \mathrm{hadrons}$, whose accuracy must
meet certain criteria
to guarantee the desired final precision
in the HVP. The fact that we are analyzing the 
anomalous moment of the muon has a direct impact on the kernel 
appearing in the dispersive integral, weighting each 
energy region differently: as a consequence  
approximately $70\%$ of the total HVP contribution to $a_\mu$
comes from the 
$\pi^+ \pi^-$ channel alone.

\section{$\tau$ input for $(g-2)_\mu$}

At a time when experimental decay rates of the $\tau$ lepton
were more precise than electron-production experiments, the
authors of \Ref{Alemany:1997tn} proposed to use the vector spectral functions
measured in hadronic $\tau$ decays to compute the HVP with 
the standard dispersive methods. 
In the following we will mostly focus our attention on the
$\pi\pi$ channel, given its importance and dominance in 
the total signal and error of $a_\mu$\footnote{Higher multiplicities
can be taken into account in principle, but lead to larger experimental
systematics.}.

The decays of the $\tau$ leptons are mediated by weak interactions 
and as consequence the observed $\pi^- \pi^0$ state is charged 
and purely isospin 1. 
Instead the hadronic vacuum polarization originates
from the electromagnetic current, thus producing intermediate
$\pi^+ \pi^-$ states which are neutral and predominantly 
isospin 1, with a small isospin-0 component as well. 
Therefore to relate the spectral functions obtained from 
$\tau$ decays, which we denote by $v_-$, to the calculation 
of the HVP, the proper isospin correction factor $\RIB$ 
is required.
The neutral spectral function $v_0$ can be
obtained according to $v_0(s) = \RIB(s) v_-(s)$, $s$ being
the invariant mass of the hadronic system, 
and isospin breaking effects contributing to the
 correction factor 

\begin{equation}
\RIB(s) = \frac{\FSR(s) }{\GEM(s) } 
\frac{\beta_0^3 (s)}{\beta_-^3(s)}
\frac{|f_0(s)|^2}{|f_-(s)|^2} 
\end{equation}
 
can be split in 
different parts\footnote{
In our list we omit short-distance electro-weak and radiative corrections, 
 accounted for by a multiplicative factor $\SEW$,
 since it is often used in the definition of $v_-$ from 
the decay-rate measurement and not in $\RIB$.}:

\begin{itemize}
\item long-distance radiative corrections, where a soft photon
is emitted from the $\tau^-$ or $\pi^-$, or exchanged between 
the two; these contributions are contained in
the function $\GEM$, computed in \Refs{Cirigliano:2001er,Cirigliano:2002pv}
in the framework of Chiral Perturbation Theory and 
in \Refs{FloresBaez:2006gf,FloresTlalpa:2006gs} with Vector Meson
Dominance models;
\item final state radiation ($\FSR$) of the $\pi^+ \pi^-$ state, 
computed assuming point-like pions as well;
\item  a factor that accounts for 
the difference in phase spaces (called $\beta$) between the $\pi^0 \pi^-$
and the $\pi^- \pi^+$ states, due to $m_{\pi_0}\neq m_{\pi^\pm}$;
\item finally the $\rho-\omega$ mixing phenomenon and differences 
in the masses and widths of the $\rho^0$ and $\rho^-$ determine
the strength of the isospin breaking 
in the two-pion charged
and neutral form factors, $f_-(s)$ and $f_0(s)$ respectively.
\end{itemize}

With this definition of the factor $\RIB$ it is then possible to compute 
the correction $\Delta a_\mu$ to be added to the anomalous magnetic moment
computed from the $\tau$ spectral functions, namely
\begin{equation}
\Delta a_\mu \equiv \int_{4 m_\pi^2}^{m_\tau^2} ds \
K(s) \big[v_0(s) - v_- (s) \big] = 
\int_{4 m_\pi^2}^{m_\tau^2} ds \
K(s) \big[ \RIB(s) - 1 \big] v_-(s) \,,
\label{eq:damu_pheno}
\end{equation}
where $K(s)$ represents the muon kernel.
Historically, three groups\footnote{
For convenience we will label the separate groups as
\emph{Cirigliano et al.}~\cite{Cirigliano:2001er,Cirigliano:2002pv},
\emph{Davier et al.}~\cite{Davier:2002dy,Davier:2003pw,Davier:2009ag} and
\emph{Jegerlehner et al.}~\cite{Jegerlehner:2011ti,Jegerlehner:2008zza}.
Note that the authors of 
\Refs{Cirigliano:2001er,Cirigliano:2002pv} 
did not include the $\FSR$.} 
have computed the isospin rotation of 
$\tau$ spectral functions
always obtaining a value for $a_\mu$ 
incompatible with the corresponding determination from $e^+ e^-$ data. 
More specifically $a_\mu[\tau]$ is bigger than $a_\mu[ee]$ by approximately 
22-25 units of $10^{-10}$, while the theoretical estimates $\Delta a_\mu$ 
were around -12, thus resulting in a systematic error for the combined 
estimate much larger than the simpler and cleaner $e^+ e^-$ determination.
For this reason and also due to dramatic improvements in the
experimental measurements of $\pi^+\pi^-$ spectral functions, in particular
in the BaBar and KLOE experiments, this alternative theoretical
determination of $a_\mu$ was abandoned, until Jegerlehner and Szafron
proposed a solution~\cite{Jegerlehner:2011ti} 
to the puzzle, by adding the effect of 
$\rho-\gamma$ mixing to the calculation of $\RIB$.

In our work we present an alternative approach to the calculation
of the correction $\Delta a_\mu$ based on Lattice QCD.
It is worth noting that the overall size of the correction, which is about 2-3 \%,
combined with the precision for $a_\mu$, $O(1 \%)$, 
results in a required uncertainty for $\Delta a_\mu$ of approximately 
10-20 \%, a goal that with current advances in Lattice QCD+QED calculations
seems achievable.

\section{Lattice calculation}

\begin{figure}[ht]
\centering
\includegraphics[width=.49\textwidth]{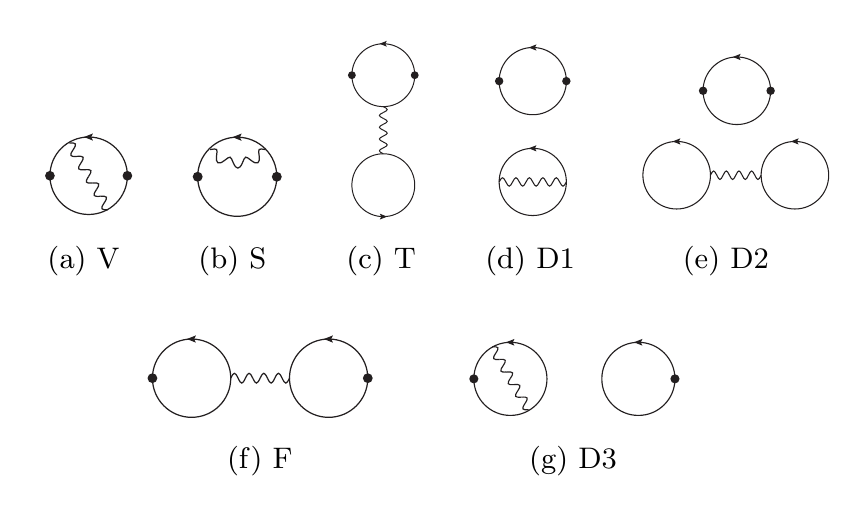}
\raisebox{2em}{\includegraphics[width=.38\textwidth]{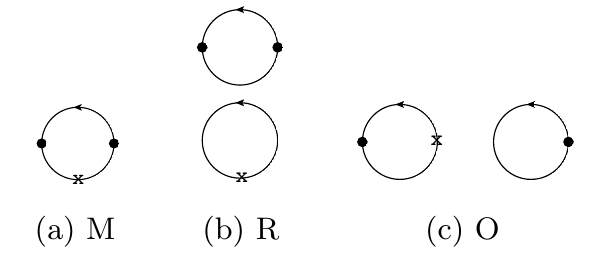}}
\caption{QED (left) and strong-isospin breaking (right) diagrams contributing
to the vector-vector two-point function. The black dots represent the EM currents
where the external photons are attached. Note the absence of tadpole diagrams, 
due to local vector currents,
for which the knowledge of the normalization factor $Z_V$ is necessary.} 
\label{fig:diagrs}
\end{figure}

The progress in the last years to include QED effects in 
Lattice QCD calculations was significant. Based on this advances,
we develop the formalism necessary to compute $\RIB$ 
on the lattice and we report some preliminary results in the next
Section before concluding.
By realizing that the electro-magnetic current is formed by
an isospin 0 and 1 parts (we consider only to the light flavors)
\begin{equation}
j_\mu^\gamma = i e \big( Q_\up \bar u \gamma_\mu u + Q_\down \bar d \gamma_\mu d \big)
= \overbrace{i e \frac{Q_\up + Q_\down}{2} 
(\bar u \gamma_\mu u + \bar d \gamma_\mu d)}^{=j_\mu^{(0)}}
+ \overbrace{i e \frac{Q_\up - Q_\down}{2} 
(\bar u \gamma_\mu u - \bar d \gamma_\mu d)}^{=j_\mu^{(1)}}
\end{equation}
we decompose the standard vector-vector two-point function in three terms
\begin{equation}
G^\gamma(t) = \frac{1}{3} \sum_{k,\,\vec x} \langle j_k^\gamma (x)  j_k^\gamma (0) \rangle = 
G_{00}^\gamma (t) + 2 G_{01}^\gamma(t) + G_{11}^\gamma(t) \,, \quad
G_{ij}^\gamma (t) \equiv \frac{1}{3} \sum_{k,\, \vec x} 
\langle j_k^{(i)} (x) j_k^{(j)} (0) \rangle \,. 
\label{eq:Gij}
\end{equation}
In \eq{Gij} the term $G^\gamma_{01}$ vanishes in the isospin limit and is
different from zero due to QED and $O(m_\up - m_\down)$ effects.
Instead both $G^\gamma_{00}$ and $G^\gamma_{11}$ survive in the isospin limit:
the former contains also disconnected contributions, while the latter
dominates the signal.
Then, to map our calculation to the $\tau^-$ spectral
functions, we compute the expectation value of two charged 
isospin 1 currents
\begin{equation}
j_\mu^{(1,-)} = i e \frac{Q_\up - Q_\down}{\sqrt 2} (\bar u \gamma_\mu d) \,, \quad
G_{11}^W \equiv \frac{1}{3} \sum_{k,\,\vec x} \,
\langle j_k^{(1,-)}(x) j_k^{(1,+)}(0) \rangle  \,.
\end{equation}
To compute the HVP contribution to $a_\mu$ we use the Time-Moment 
representation~\cite{Bernecker:2011gh}, 
$a_\mu \equiv 4 \alpha^2 \sum_t w_t G^\gamma(t) $,
and following the decomposition in \eq{Gij}
we obtain correspondingly three contributions to $a_\mu$. 
At this point is it straightforward to define the correction
$\Delta a_\mu$ in our lattice calculation
\begin{equation}
\Delta a_\mu \equiv 4 \alpha^2 \sum_t w_t \big[ G^\gamma(t) - G^W(t) \big] \quad \to \quad
\Delta a_\mu[\pi\pi] = 4 \alpha^2 \sum_t w_t \big[ 2 G^\gamma_{01}(t) + 
\overbrace{G^\gamma_{11}(t) - G^W_{11} (t)}^{=\delta G(t)} \big] \,,
\label{eq:damu_lat}
\end{equation}
where in the second equation we have neglected the $G^\gamma_{00}$ term
that contributes only to channels with an odd number of pions.
Also in this case, in the isospin limit $\Delta a_\mu$ goes to zero; 
therefore to compute this quantity we perform a diagrammatic expansion
where we insert a single photon in all possible ways, together with 
all possible insertions of the scalar operator. In \fig{diagrs} we 
draw the leading contributions and we postpone the discussion on the
details of the QED formalism on the lattice to the next section.
For the reader's convenience we report below, as an example, 
the first diagram\footnote{We keep minus signs from Dirac traces, 
photon coupling, etc.. all explicit.}
in \fig{diagrs} which we call $V$
\begin{equation}
V = \frac{1}{3} \sum_{k, \vec x} \mathrm{Tr} \ 
\langle \Big[ \sum_{z,y} \sum_{\mu\nu}
\gamma_k D^{-1} (x,z) \gamma_\mu D^{-1}(z,0) \,
\gamma_k D^{-1} (0,y) \gamma_\nu D^{-1}(y,x) \cdot \Delta^{\mu\nu}(z,y)
\Big] \rangle \,,
\end{equation}
with $D^{-1}$ being a (light) quark propagator and $\Delta^{\mu\nu}(z,y)$ the photon
propagator\footnote{In this work we have used the Feynman gauge.}.

Studying separately the pure $I=1$ correction $\delta G$ has the advantage
that many terms simplify and it can be directly mapped to the 
mass and width differences of the neutral and charged $\rho$ mesons
in the phenomenological models of the form factors. 
For this quantity we obtain\footnote{
The factor $Z_V^2$ associated to the external photons is 
included in our definition of the weights $w_t$.}
\begin{equation}
\delta G \equiv G^\gamma_{11} - G^W_{11} = 
 - (4 \pi \alpha) Z_V^2 \frac{(Q_\up - Q_\down)^4}{4} 
\big[ V - F \big] \,.
\label{eq:deltaG}
\end{equation}
Instead, in the contribution originating from the isospin 0 to 1 transition,
which from the phenomenological point of view 
is very interesting due to the $\rho-\omega$ mixing,
cancellations are less relevant, resulting in a dependence on many
diagrams:
\begin{equation}
2 G^\gamma_{01} = - \frac{(Q_\up^2 - Q_\down^2)^2}{2} (4 \pi \alpha) 
Z_V^2 \big[V + 2S - F + \cdots\big] - 
\frac{Q_\up^2-Q_\down^2}{2} (m_\up - m_\down) \big[2M - 2O + \cdots\big] \,.
\end{equation}
We note that as expected these isospin breaking functions are 
proportional to $(Q_\up - Q_\down)$ and $(m_\up - m_\down)$.\\

Our final discussion point before turning to the numerical results, 
is on the connection between the previous theoretical determinations of 
$\Delta a_\mu$ and ours. The integral of $\Delta a_\mu$ can be compared 
without problems, in the continuum and infinite volume limit.
However given the large amount of cancellations in the integrands in 
\eq{damu_pheno} and in \eq{damu_lat}, it may be worth exploring the possibility
of doing a more detailed comparison between the two determinations.
To achieve this, first we realize that the lattice calculation is 
performed in Euclidean time and can not be analytically continued to 
Minkowski space. Therefore it is natural to apply a simple Laplace
transform and convert the standard determinations of $\RIB(s)$ 
to Euclidean time
\begin{equation}
\Delta a_\mu[\pi\pi,\tau] = 4 \alpha^2 \sum_t w_t \times \Big\lbrace
\frac{1}{12 \pi^2} \int d \omega \ \omega^2 e^{-\omega t}
\big[ \RIB(\omega^2) -1 \big] v_-(\omega^2) \Big\rbrace \,.
\end{equation}

\section{Preliminary results}

In our work we consider
the results published by the RBC/UKQCD collaboration on the 
calculation of $a_\mu$, including the leading QED and 
strong-isospin diagrams~\cite{Blum:2018mom}.
We perform a simple re-analysis of these results to compute the
quantities of interest for this work.\\

Firstly, when including QED in lattice calculations a precise
prescription to remove the zero-modes is required 
and in our work we adopt the QED$_\mathrm L$ formalism.
Moreover we consider a diagrammatic
expansion at $O(\alphaem)$ and $O(m_\up - m_\down)$ which completely
defines our scheme: hence in the renormalization procedure,
a new set of (hadronic)
quantities must be computed (at $O(\alphaem)$ and $O(m_\up - m_\down)$) 
to tune the bare parameters of our calculation
to follow a line of constant physics.
In \Ref{Blum:2018mom} the mass of the Omega baryon $\Omega^-$ has been used
to re-compute the lattice spacing, while the mass difference $m_{\pi_0} - m_{\pi^+}$
and the charged pion mass were used to define $m_\up$ and $m_\down$ separately\footnote{
In \Ref{Blum:2018mom} also the strange quark mass has been properly retuned, 
but this does not affect our work since we consider only up-down valence 
contributions to $a_\mu$.}.
Similarly $Z_V$ has been recomputed to include $O(\alphaem)$ effects~\cite{Blum:2018mom}.

\begin{figure}[ht]
\includegraphics[width=.49\textwidth,trim={0 0 0 1.1cm},clip]{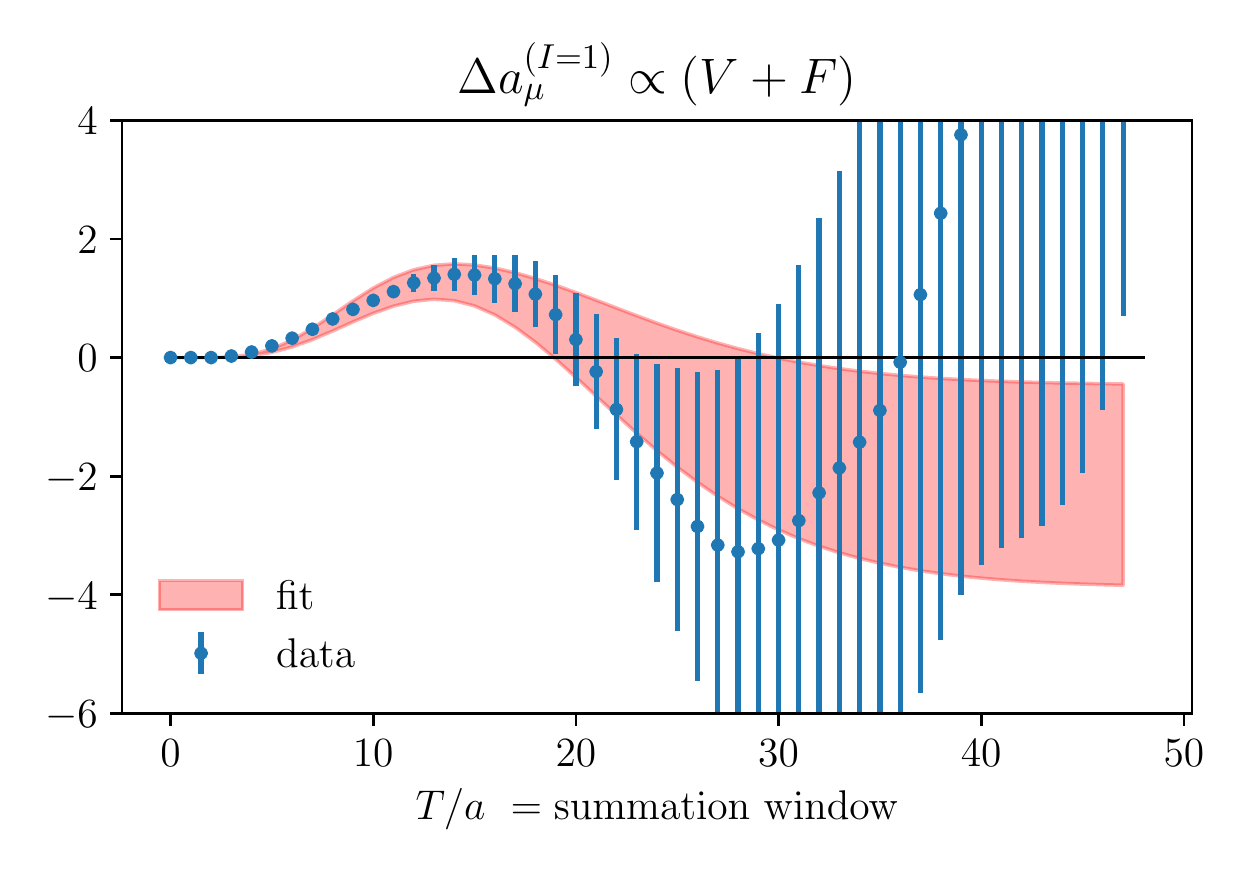}
\includegraphics[width=.49\textwidth,trim={0 0 0 1.1cm},clip]{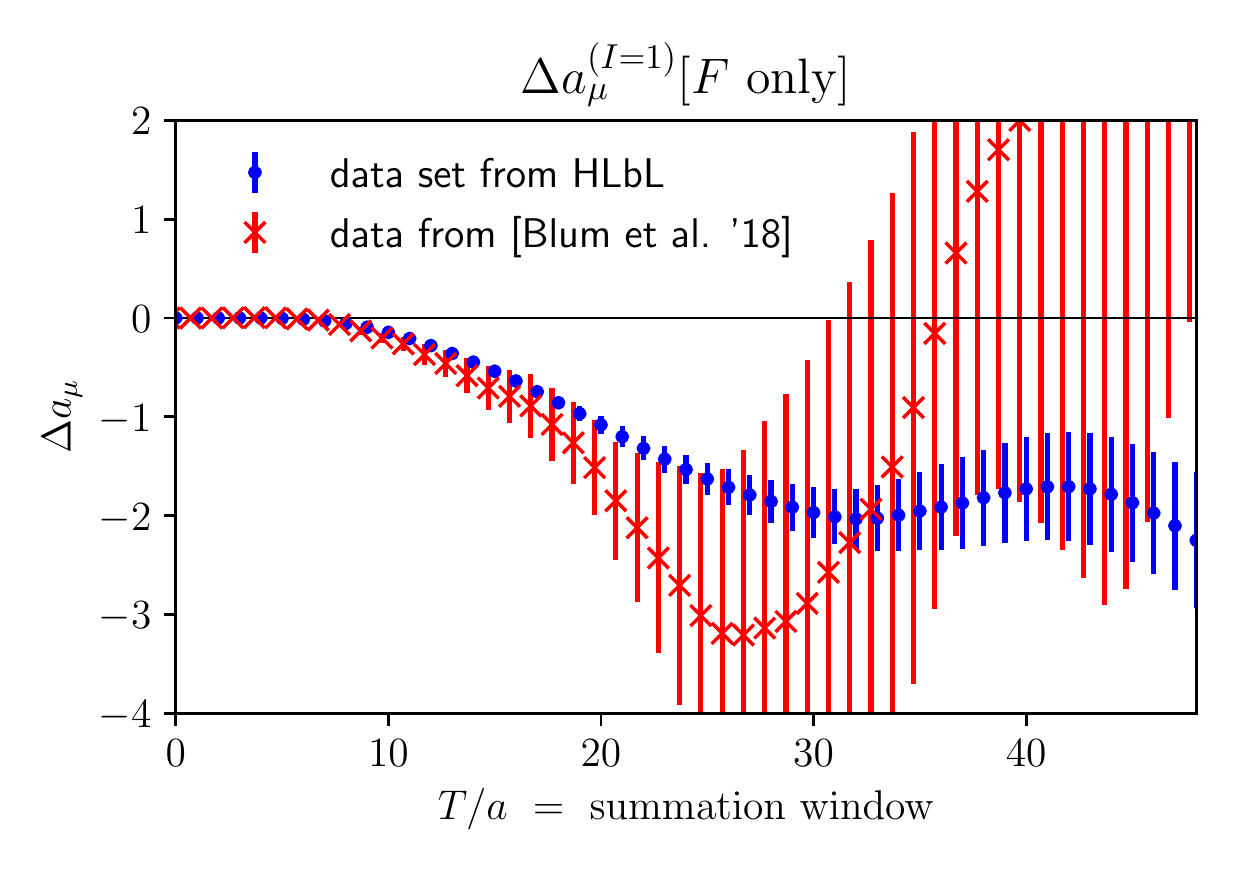}
\caption{{\it Left:} pure $I=1$ part of $\Delta a_\mu$ ($\delta G$ only)
as a function of the 
summation window in lattice units. The fit to the lowest state with constrained energy
provides a substantial reduction in the statistical noise. We estimate the 
systematic error by varying the energy between the two-pion and pion-photon 
states.
{\it Right:} contribution of diagram $F$ to the pure $I=1$ part of $\Delta a_\mu$ 
(i.e. contribution of diagram $F$ to $\delta G$).
The red crosses correspond to the data set published in \Ref{Blum:2018mom} 
and used in the rest of this work. The blue data points correspond to a new ongoing
re-analysis of the data produced to study the HLbL contribution to $a_\mu$~\cite{Blum:2016lnc},
which provides a significant statistical improvement for this observable.}
\label{fig:syst}
\end{figure}

\begin{figure}[ht]
\centering
\includegraphics[width=.49\textwidth]{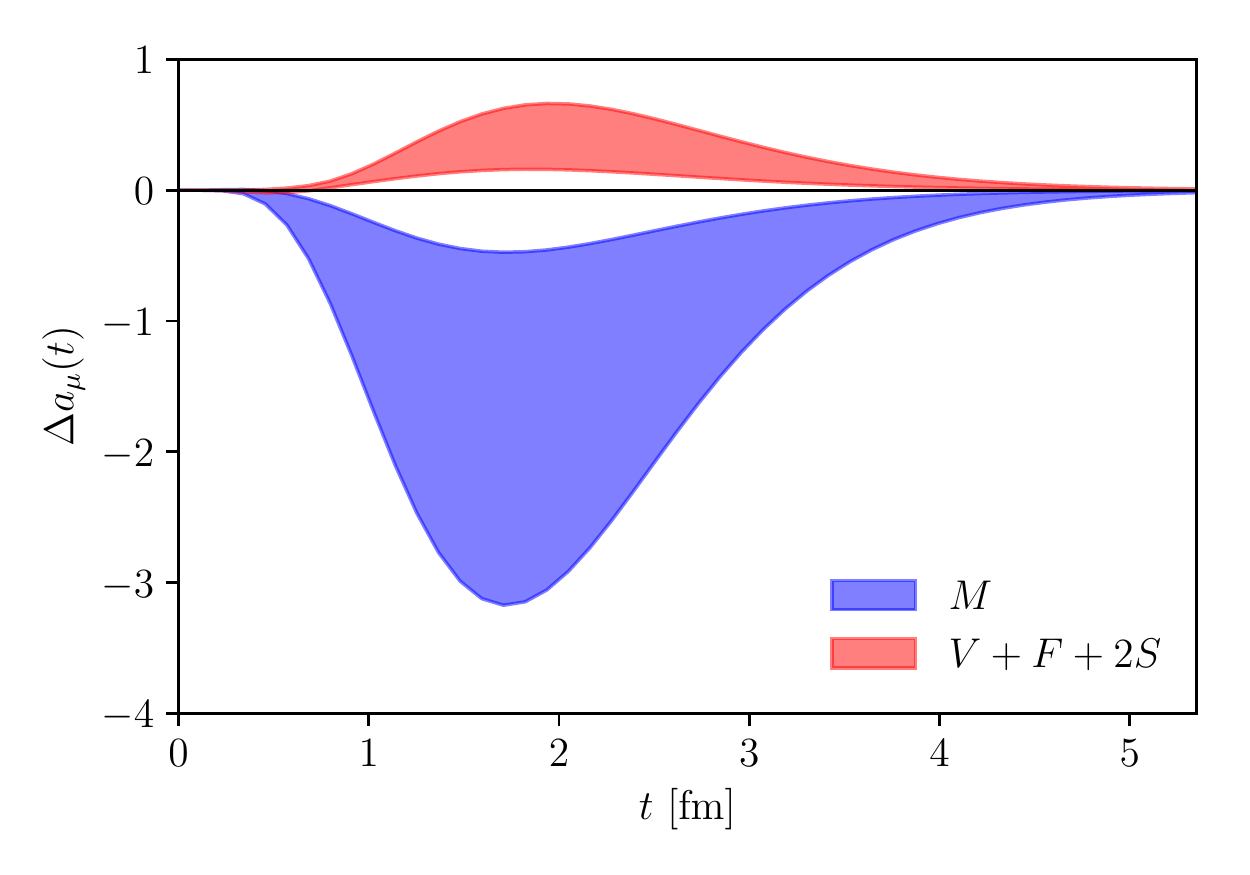}
\includegraphics[width=.49\textwidth]{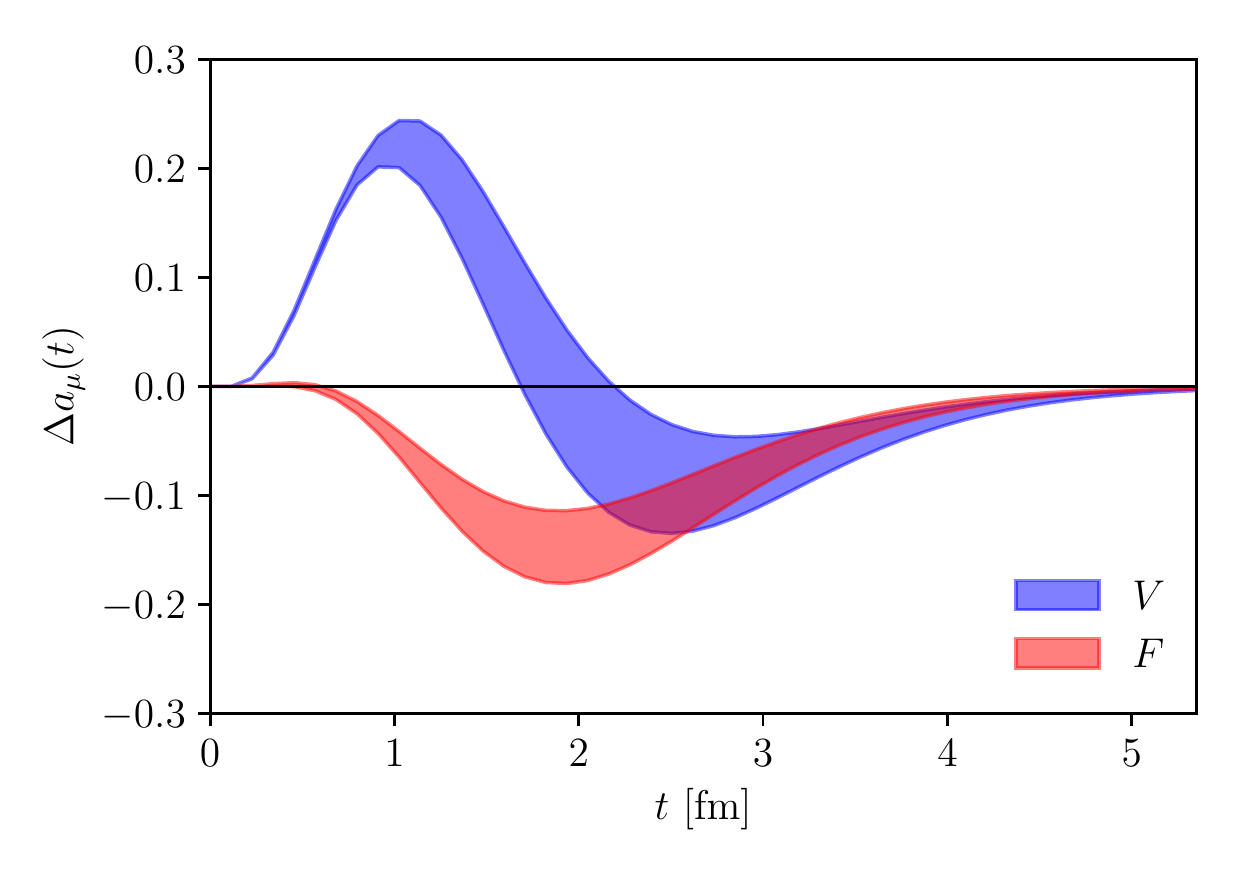}
\caption{{\it Left:} contribution to $\Delta a_\mu$ proportional to $G_{01}^\gamma$
where both QED and strong-isospin are present. 
Unfortunately the present quality of the connected strong-isospin term ($M$) 
is sufficient to bound the value of $\Delta a_\mu$ but is practically compatible
with zero for all Euclidean times. For this reason we will invest additional
resources to improve this specific part and we do not proceed much
further now. Note also that sub-leading diagrams suppressed by 
at least one power of $1/N_c$ or $1/N_f$ are neglected.
{\it Right:} pure $I=1$ term of $\Delta a_\mu$. In this case all diagrams are included
and only QED contributes, simplifying the calculation. A relatively good signal is 
already achieved with current statistics, highlighting the high degree of cancellations
present when computing the integral. For this reason we stress again the importance 
of a more detailed comparison of the integrands between the lattice and previous phenomenological
determinations.
}
\label{fig:results}
\end{figure}

In \fig{results} we show our preliminary results obtained at fixed lattice spacing
($a^{-1} \simeq 1.73 ~\mathrm{GeV}$) and fixed volume ($L\approx ~5.4 \mathrm{fm}$).
The calculation is performed at physical pion masses
and we refer the reader to \Ref{Blum:2018mom} for more details on the numerical
strategies used to compute every diagram.
To improve the statistical error of our estimates we rely on a constrained fit
with functional form $(c_0 + c_1 t) e^{-Et}$, also discussed in \Ref{Blum:2018mom},
where the energy is fixed to be either $E_{\pi\pi}$ or $E_{\pi \gamma}$, which
are very similar due to the specific choice of $L$. The systematic
error arising from the difference of the two fits is still much smaller than
the statistical one, and the fact that the energy is constrained reduces the 
noise in the long tails. We demonstrate this effect in the left panel of \fig{syst}
where we show the partial sum $\sum_t^T w_t \delta G(t)$ as a function of the summation window.
In order to improve this calculation we are also working
on a new and better determination of these diagrams where such fits can be 
pushed to much later (Euclidean) times, thus significantly reducing the systematics.
In the right panel of \fig{syst} we present the comparison between the current and
the new determination of diagram $F$, obtained from a re-analysis of the point-source
propagators generated to study the Hadronic Light-by-Light contribution to 
$a_\mu$~\cite{Blum:2016lnc}.

In the present exploratory study our goal is to understand the size of the various
terms to better plan these future calculations, where we will significantly reduce
the dependence of our results on these assumptions.
As we can see from the comparison of the two panels of \fig{results},
the pure $I=1$ contribution $\delta G$, where we include all diagrams (see \eq{deltaG}) 
and only QED matters at this order,
has little impact compared to the isospin 0-1 channel, 
which in turn is completely dominated by strong-isospin breaking. This expectation
matches ChPT predictions for the $\rho-\omega$ mixing parameter, where $O(m_\up-m_\down)$
terms are expected to be a factor 4 larger than QED.
However, we must note that the current determination of the connected SIB term, namely 
diagram $M$ in our notation, is unable to resolve it from
zero throughout the entire time extent.
For this reason, with the current data we can not provide a reliable estimate of $\Delta a_\mu$.
Nevertheless, in the near future we expect to be able to significantly
improve our estimates of all the diagrams presented in this study, together with 
the sub-leading ones.
Due to the presence of QED interactions and SIB we expect in general large finite volume effects.
Part of our efforts are in fact devoted to a careful study of these systematic uncertainties
by repeating our calculation on several lattices, differing only by the volume, 
together with the insights provided by ChPT. 

\section{Conclusions}

In this work we have presented some preliminary results for the isospin-breaking 
corrections necessary to compute $a_\mu$ from $\tau$ decays. 
Lattice QCD+QED calculations are finally mature enough to attack this class of problems
and we have showed how a simple re-analysis of already published results can 
give further insights into this quantity.
At first we have developed the formalism necessary to define the quantity of interest
($\Delta a_\mu$) on the lattice and with the current available data we have attempted
a first calculation. 
We have demonstrated that only two QED diagrams are needed to compute the 
the pure $I=1$ component of the isospin-breaking rotation and we have provided some
initial numerical evidence. Systematic errors such as finite volume and discretization
effects have not been estimated and we refer the reader to future publications, where
we plan to study both. We have also demonstrated our current progress in the calculation
by showing a new determination of diagram $F$, which is substantially more precise, 
allowing us to reduce some systematics introduced by our fitting procedure.
Finally we stress once more the importance of the comparison of the integrands, 
beyond the final results for $\Delta a_\mu$, due to the high level of cancellations
taking place. We are working together with the other groups that computed this
quantity in the past towards a careful comparison between our Lattice QCD+QED results and 
the previous theoretical/phenomenological determinations. With more and exciting 
experimental results ahead of us an alternative, competitive and solid estimate of $a_\mu$ 
from $\tau$ decays may again be possible.

\section*{Acknowledgements}
We would like to thank our RBC/UKQCD collaborators for helpful discussions and support.
M.B. is particularly indebeted to F.~Jegerlehner for sharing private results, 
to K.~Maltam for helpful insights on $\tau$ phenomenology 
and to M.~Davier for stimulating discussions.
This work is supported in part by US DOE Contract DESC0012704(BNL). 
T.I. is supported by JSPS KAKENHI grant numbers JP17H02906.
C.L. is also supported by a DOE Office of Science Early Career Award.

\bibliographystyle{JHEP}
\bibliography{lat18.bib}

\end{document}